\documentclass[prx,twocolumn,showpacs,superscriptaddress,preprintnumbers,amssymb]{revtex4-2} 
\usepackage{graphicx}
\usepackage{latexsym}
\usepackage{amssymb}
\usepackage{amsmath}
\usepackage{mathtools}
\usepackage{amsmath} 
\usepackage{amsfonts} 
\usepackage{upgreek}
\usepackage{bm}
\usepackage{multirow}
\usepackage{enumitem}
\usepackage{color}
\usepackage[colorlinks, citecolor=blue]{hyperref}
\usepackage{physics}
\usepackage{tcolorbox}
\usepackage{manfnt}
\usepackage{lipsum}
\usepackage[ruled,linesnumbered]{algorithm2e}
\usepackage{tabularx}
\usepackage{nicematrix}
\usepackage{braket}
\usepackage{bm}
\usepackage[normalem]{ulem}

\NiceMatrixOptions{
code-for-first-row = \color{blue} ,
code-for-last-row = \color{blue} ,
code-for-first-col = \color{blue} ,
code-for-last-col = \color{blue}
}

\usepackage{amsthm}
\theoremstyle{definition}

\newtheorem{definition}{Definition}

\newcommand{\J}{\mathcal{J}}
\newcommand{\SYK}{{\rm SYK}}
\newcommand{\TFD}{{\rm TFD}}
\newcommand{\hop}{{\rm s}}
\newcommand{\inte}{{\rm d}}

\newcommand{\both}{{\rm both}}
\newcommand{\bcS}{{\bm{\cS}}}
\newcommand{\err}{{\rm err}}
\newcommand{\thr}{{\rm th}}

\newcommand{\gam}{{\bm{\gamma}}}

\newcommand{\sgn}{{\rm sgn}}

\newcommand{\cO}{ {\cal O} }
\newcommand{\cD}{ {\cal D} }

\newcommand{\cH}{ {\cal H} }

\newcommand{\cS}{ {\cal S} }

\newcommand{\BigO}{{\mathcal{O}}}

\newcommand{\appref}[1]{SI.\,\ref{#1}}
\newcommand{\eqnref}[1]{Eq.\,\eqref{#1}}

\newcommand{\figref}[1]{Fig.\,\ref{#1}}

\newcommand{\cN}{{\mathcal{N}}}
\newcommand{\im}{\mathbf{i}}
\newcommand{\interp}{{q}}

\newcommand{\rAngle}{\rangle \hspace{-2pt} \rangle }
\newcommand{\lAngle}{\langle \hspace{-2pt} \langle }
 
\renewcommand{\tr}{\mathrm{tr}}

\newcounter{protocol}

\begin{document}

\title{Error Threshold of SYK Codes from Strong-to-Weak Parity Symmetry Breaking}

\author{Jaewon Kim}
\email{jaewonkim@berkeley.edu}
\affiliation{Department of Physics, University of California, Berkeley, CA 94720, USA}

\author{Ehud Altman}
\email{ehud.altman@berkeley.edu}
\affiliation{Department of Physics, University of California, Berkeley, CA 94720, USA}
\affiliation{Materials Sciences Division, Lawrence Berkeley National Laboratory, Berkeley, CA 94720, USA}

\author{Jong Yeon Lee}
\email{jongyeon@illinois.edu}
\affiliation{Physics Department, University of Illinois at Urbana-Champaign, Urbana, Illinois 61801, USA}

\date{\today}
\begin{abstract}
Quantum error correction (QEC) codes are fundamentally linked to quantum phases of matter: 
the degenerate ground state manifold corresponds to the code space, while topological excitations represent error syndromes. Building on this concept, the Sachdev-Ye-Kitaev (SYK) model, characterized by its extensive quasi-ground state degeneracy, serves as a constant rate approximate QEC code. 
In this work, we study the impacts of decoherence on the information-theoretic capacity of SYK models and their variants. Such a capacity is closely tied to traversable wormholes via its thermofield double state, which theoretically enables the teleportation of information across a black hole. We calculate the coherent information in the maximally entangled quasi-ground state space of the SYK models under the fermion parity breaking and parity conserving noise.  Interestingly, we find that under the strong fermion parity symmetric noise, the mixed state undergoes the strong to weak spontaneous symmetry breaking of fermion parity, which also corresponds to the information-theoretic transition. 
Our results highlight the degradation of wormhole traversability in realistic quantum scenarios, as well as providing critical insights into the behavior of approximate constant-rate QEC codes under decoherence.
\end{abstract} 
\maketitle

\emph{Introduction.}---
Protecting quantum information from noise is a central challenge in quantum computing, often addressed by encoding logical qubits into a larger set of physical qubits and employing error correction procedures. Recent advancements include the discovery of good quantum low-density parity-check (qLDPC) codes~\cite{panteleev2022asymptotically, dinur_qldpc_decoder}, where both the number of logical qubits $k$ and the distance $d$ scale with the system size $N$, while each qubit interacts with a finite number of error-detecting check operators.

An important property of the quantum codes that is yet to be better understood is their error threshold--the maximum error rate below which error correction is possible. It is natural to explore this using intrinsic information-theoretic metrics, such as the quantum channel capacity, as done recently for a broad family of quantum states and CSS codes~\cite{fan2023diagnostics, Lee2023PRXQ, bao2023mixedstate, chen2024separability, lee2024exact, niwa2024coherentinformationcsscodes}. However, in this case, such a calculation may be hindered by the complex geometry of the interaction graph. While the extensive scaling of the code distance suggests a finite error threshold, identifying its location or the nature of the transition requires better conceptual insights. For instance, it remains unclear whether the code capacity drops abruptly to zero or gradually diminishes beyond the threshold.

\begin{figure}[!t]
    \centering
    \includegraphics[width = 0.98\columnwidth]{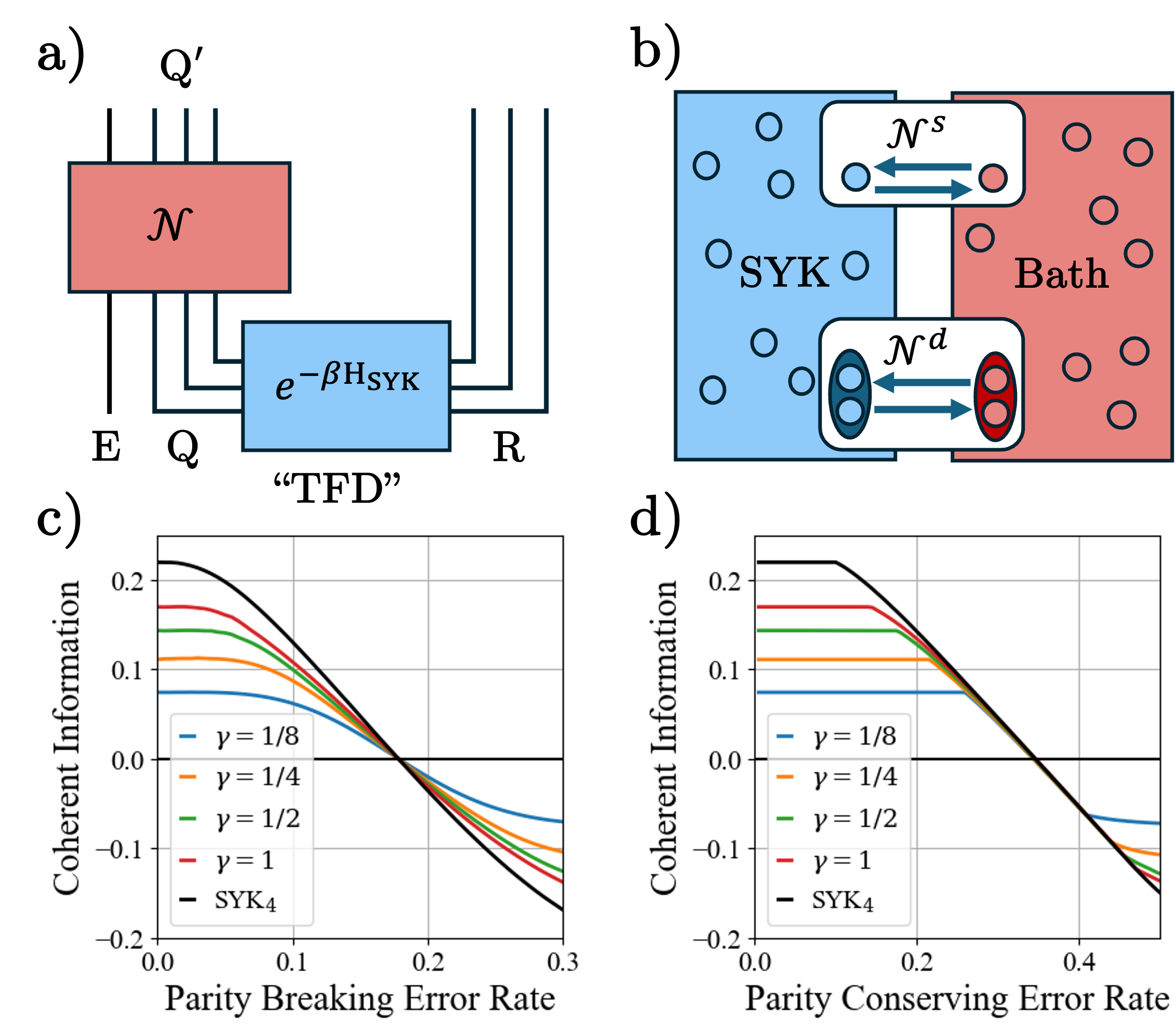}
    \caption{{\bf Summary.} (a) Circuit diagram of our set-up. The system ($Q$) is prepared in a thermofield double state with the reference ($R$), and undergoes decoherence ($\cN$) into $Q'$. (b) The types of noise that we consider. Error $\cN^{\hop}$ exchanges a single majorana with the bath and breaks fermion parity. Error $\cN^{\inte}$ exchanges two majoranas with the bath and conserves fermion parity. (c,d) The Renyi-2 coherent information as a function of the error rate for parity breaking errors (c) and parity conserving errors (d) for the SYK model (black line) and low-rank SYK model at various ranks (colored lines). We see no threshold behavior for parity breaking errors but in contrast find a robust error threshold for parity conserving errors.}
    \label{fig:Main}
\end{figure}

Motivated by these challenges, we study the error threshold in Sachdev-Ye-Kitaev (SYK) models~\cite{Sachdev_1993,KitaevTalk, Maldacena_2016, maldacena2016conformal} and their low-rank variants~\cite{lowrank}, which approximate certain aspects of good qLDPC codes while offering more analytic control~\cite{Chandrasekaran_2022, bentsen2023approximate, yi2023complexity}.
Notably, SYK models exhibit a non-vanishing entropy density in the zero-temperature limit, suggesting that their low-energy states can be regarded as the logical subspace of an approximate quantum error-correcting code (QEC) with the number of logical qubits proportional to $N$.
% The fact that these models exhibit non-vanishing entropy density in the zero temperature limit suggests that their low energy states may be regarded as the logical subspace of an approximate QEC with the number of logical qubits proportional to $N$. 
Furthermore, Ref.~\cite{bentsen2023approximate} estimated the code distance as $d\sim N^{2\Delta}$, where $\Delta$ is the scaling dimension of a fermion. In the standard SYK model with $q$-fermion interactions $\Delta\,{=}\,1/q$, but in low-rank SYK models the scaling dimension can be varied continuously in the range $1/4\,{<}\,\Delta\,{<}\,1/2$ by tuning the rank of the four-fermion interaction matrix. Thus, low-rank SYK models can be tuned arbitrarily close to a ``good'' code (albeit not low density due to all-to-all interactions).
However, the robustness to physical errors, specifically in terms of establishing an error threshold, was not established. The question of the error threshold in SYK models may also be relevant to the robustness of quantum protocols, such as traversable wormhole teleportation, which rely on this system's special information scrambling and encoding properties~\cite{Maldacena_2017, gao2021traversablewormholeteleportationprotocol}.

To identify the threshold, we calculate the coherent information of the channel associated with the thermofield double of the generalized SYK models subjected to noise.
This coherent information provides a strict upper bound on the amount of decodable information~\cite{coherentInfo1, coherentInfo2}.
Through numerical and perturbative studies, we find that when the noise channel conserves fermion parity as a strong symmetry~\cite{Bu_a_2012, PhysRevA.89.022118, de_Groot_2022}, the coherent information of SYK-type models undergoes a decoherence induced transition~\cite{lee2022symmetry, fan2023diagnostics, Lee2023PRXQ, bao2023mixedstate, chen2024separability, chen2023symmetryenforced, wang2023intrinsic, guo2023twodimensional, sang2023mixedstate, colmenarez2023accurate, su2024tapestry, lyons2024understanding,
lee2024exact,
li2024replicatopologicalorderquantum,
chen2024unconventionaltopologicalmixedstatetransition, niwa2024coherentinformationcsscodes}, signifying a threshold behavior of decodable information. In particular, this information-theoretic transition is tied to spontaneous breaking of the fermion parity from a strong to weak symmetry~\cite{Lee2023PRXQ, StW_U(1), sala2024spontaneousstrongsymmetrybreaking, lessa2024strongtoweakspontaneoussymmetrybreaking,gu2024spontaneoussymmetrybreakingopen}.
Our work provides critical insights into the behavior of approximate QEC codes under decoherence, with implications for the potential performance of constant-rate quantum error-correcting codes.

\emph{Model.}---The family of SYK models we consider consists of $N$ Majorana fermions with all-to-all interactions:
\begin{equation}
    H_\SYK = \sum_{ijkl=1}^N J_{ijkl} \gamma^i \gamma^j \gamma^k \gamma^l.
    \label{eq:SYK}
\end{equation}
In the standard SYK model, $J_{ijkl}$ is a generic real antisymmetric tensor with zero mean and variance equal to $\J^2/6N^3$. In this case, by grouping the indices into pairs, the interaction tensor can be viewed as a full rank $N^2\,{\times}\,N^2$ matrix. We also consider a generalized family of models, known as \emph{low-rank} SYK models, in which the interaction tensor has rank $N$~\cite{lowrank}. Specifically we take $J_{ij,kl}\,{=}\,\sum_{n = 1}^{\gam N} u_{ij}^n u_{kl}^n$. Here $\{ u^n \}_{n=1}^{\gam N}$ are real random matrices of zero mean and covariance $\overline{u_{ij}^n u_{kl}^m}\,{=}\, (g/N)^2 \delta_{ik} \delta_{jl} \delta_{nm}$ and $\gam$ is a parameter that continuously tunes the rank of the interaction matrix.

Both the standard and the low-rank SYK models realize a conformal critical state at low energies but with different scaling dimensions $\Delta$ of the fermion operator. In the standard SYK model $\Delta\,{=}\,1/4$, while in the low-rank models $\Delta$ can be varied continuously in the range $(1/2,1/4)$ by tuning the rank parameter $\gam$.  
Crucially, both the standard and low-rank SYK model have an extensive zero-temperature entropy $S_0 = s_0(\gam) N$, where $s_0(\gam)$ is an increasing function of $\gam$ (See \appref{app:SYK}).

The zero temperature entropy implies an extensive quasi-degeneracy of the ground state that can be used to encode ${\sim}Ns_0 $ logical qubits. Thus the SYK models can be regarded as QEC with $\BigO(1)$ code rate.
Note, however, that because this system is gapless it can only form an approximate QEC code~\cite{yi2023complexity, sang2024approximatequantumerrorcorrecting}, where error recovery cannot be exact.

One may also characterize the code distance~\cite{KnillLaflamme} of the SYK models through the generalization of this notion to approximate QEC. To this end, the $\epsilon$-distance is defined as the minimum size of an error operator that can change the logical information by $\epsilon$~\cite{yi2023complexity}. With this notion, the code distance in the SYK codes scales as $N^{2\Delta}$~\cite{bentsen2023approximate}, where $\Delta$ is the fermion scaling dimension. Interestingly, for $\gam\to 0$ we have $\Delta\to 1/2$ and therefore the code distance approaches $N$. This means that we can approach arbitrarily close to a ``good code'', but not quite get there, because the distance scales subextensively for any $\gam\,{>}\,0$.

An important open question is whether SYK models, viewed as QECs, have a finite error threshold. This is not obvious because these systems are gapless, and the code distance scales sub-extensively. To explore this, we analyze the resilience of the encoded quantum information under two decoherence channels: one that preserves fermion parity and another that breaks it. 
We find that the coherent information from the encoded input to the output of the channel continuously degrades under parity-breaking noise, whereas it remains robust against parity-conserving noise up to a sharp threshold, beyond which it decreases continuously with increasing error rates.
It is worth noting that the robustness of coherent information does not necessarily imply an efficient method for encoding and decoding. In fact, for the SYK model, no such schemes are currently known.

\emph{Coherent information}---
To quantify the amount of logical information that can be stored robustly in the quasi-groundstate manifold, we introduce the thermofield double (TFD) state between the system $Q$ and the reference $R$ as in \figref{fig:Main}a) at inverse temperature $\beta$:
\begin{equation}
    \ket{\textrm{TFD}} = Z_{\beta}^{-1/2} e^{-\beta H_R/2} \ket{\Phi_{QR}}
    \label{eq:H_TFD}
\end{equation}
where $\ket{\Phi_{QR}}$ denotes the maximally entangled fermionic state between the system and the reference; it satisfies the condition $(\gamma_Q^j\,{-}\,\im \gamma_R^j)\ket{\Phi_{QR}}\,{=}\,0$.
Here $\gamma^j_Q$ and $\gamma^j_R$ are Majorana fermions of the system $Q$ and the reference $R$ respectively.
$H_R$ is a SYK-type Hamiltonian as in \eqnref{eq:SYK} acting only on $R$ and $Z_\beta\,{=}\,\Tr e^{-\beta H_R}$. The thermofield double should be understood through the Choi-Jamiolkowski isomorphism~\cite{CHOI1975, JAMIOLKOWSKI1972} as the state representation of the encoding map from an input $R$ to the degrees of freedom making up the SYK model.

In the limit $N\,{\rightarrow}\,\infty$ followed by $T\,{\rightarrow}\, 0$, the entanglement entropy $S(Q)\,{=}\,S(R)\,{=}\,s_0 N$ quantifies the number of logical qubits $k$ encoded in the quasi-degenerate low-energy manifold. The order of limits is crucial because for any finite $N$ there is a unique ground state and therefore no encoding.
We enforce the order of limits by taking $T/J\,{\sim}\,N^{-\alpha}$ with $\alpha\,{<}\,1$.

More generally, the amount of quantum information that can be faithfully transmitted from the input $R$ to the output $Q'$ over a noisy channel $\cN$ is given by the coherent information (See \figref{fig:Main}a), 
\begin{align}
    I_c(Q;\cN) := S_{Q'} - S_{Q'R},
\end{align}
This quantity thus provides a fundamental upper-bound on the error correction threshold~\cite{coherentInfo1, coherentInfo2}. In absence of decoherence $S_{QR}\,{=}\,0$ and therefore $I_c(Q)\,{:=}\,S_{Q}\,{=}\,S_{R}$. The data processing inequality, together with strong subadditivity implies that the coherent information of the decohered system obeys the following inequality $-I_c(Q) \leq I_c(Q;\cN) \leq I_c(Q)$.

In usual QEC, the error threshold (for code capacity) is defined as the maximum error rate where $I_c(Q;\cN)\,{=}\,I_c(Q)$ in the thermodynamic limit~\footnote{Or more precisely, $I_c(Q;\cN)$ can be made arbitrarily close to $I^{(0)}_c(Q)$ by increasing the system size.}. In approximate QEC codes, it is often necessary to define a threshold tolerating a small fractional loss of logical qubits. In particular, the $\epsilon$-threshold is defined as follows:
\begin{definition}[$\epsilon$-threshold]
    Consider an error channel $\cN_p$ parametrized by the error rate $p$. The $\epsilon$-error threshold against $\cN_p$ is defined as
    \begin{align}
        p_{\thr}(\epsilon:\cN) := \textrm{argmax}_p \Big[\, I_c(Q;\cN_p) > (1-\epsilon) I_c(Q) \, \Big]
    \end{align}
\end{definition}
For a constant rate code, this means that a $\epsilon\,{=}\,O(1)$ threshold tolerates a loss of a finite fraction, i.e. extensive number of logical qubits; on the other hand, $\epsilon\,{=}\,1/N$ threshold tolerates only a loss of a constant $\BigO(1)$ number of qubits. 
With this definition, we consider the two classes of error channels operating on the SYK models:
\begin{subequations}
\label{eq:cN}
    \begin{align}
    & \cN^\hop_p = \prod_i \cN^\hop_{p,i}, \,\, \cN^\hop_{p,i}[\rho] = (1-p) \rho + 2p \gamma_i \rho \gamma_i \\
    & \cN^\inte_{\interp} = \prod_{ij} \cN^\inte_{\interp,ij}, \,\, \cN^\inte_{\interp,ij}[\rho] = \left(1-\frac{\interp}{N} \right) \rho + \frac{4\interp}{N} \gamma_i \gamma_j \rho \gamma_j \gamma_i\,.
    \end{align}
\end{subequations}
The Kraus operators of the channel $\cN^\hop_p$ break the fermion parity, while those of $\cN^\inte_{\interp}$ preserve this symmetry.
Equivalently, we may say that the channel $\cN_p^\hop$ preserves only weak fermion parity symmetry, while $\cN^\inte_{\interp}$ preserves the strong fermion parity symmetry.
Note that the strength of each parity conserving $\cN^\inte_{ij}$ needs to scale as $1/N$ to ensure an overall extensive effect because the number of participating pairs in the parity conserving noise is $\BigO(N^2)$.

To understand the error resilience of the SYK codes, we calculate the coherent information under $\cN$. To facilitate its calculations we define, as usual, the Renyi-version of the coherent information:
\begin{equation}
    I^{(n)}_c(Q;\cN) := S_{Q'}^{(n)}   - S_{Q'R}^{(n)}.
    \label{eq:Renyi2CI}
\end{equation}
where $S^{(n)}(\rho)\,{:=}\,\frac{1}{1-n} \log \tr \rho^n$. The von Neumann entanglement entropy and coherent information can be found through the extrapolation $\lim_{n \rightarrow 1} S^{(n)}\,{=}\,S$.

In the calculation of Renyi-$n$ quantities, it is convenient to use the Choi-Jamiolkowski isomorphism~\cite{CHOI1975, JAMIOLKOWSKI1972}; under this isomorphism, a density matrix maps to a pure state, and a superoperator (channel) maps to an operator in the doubled Hilbert space $\cH \otimes \bar{\cH}$.
The mapping is defined as $ \Vert \rho \rAngle = (\rho \otimes I) \Vert \Phi \rAngle$ where $\Vert \Phi \rAngle$ is the maximally entangled state between $\cH$ and $\bar{\cH}$. Since $\cH = \cH_R \otimes \cH_Q$ with $|\cH_R|\,{=}\,|\cH_Q|$, 
we can define two different types of maximally entangled states: $(i)$
$\Vert \Phi \rAngle := |\Phi_{R \bar{R}} \rangle | \Phi_{Q \bar{Q}} \rangle$, 
and $(ii)$ $\Vert \Psi \rAngle := \ket{\Phi_{\vphantom{\bar{Q}}QR}} \ket{\Phi_{\bar{R} \bar{Q}}}$
~\footnote{Here, $\ket{\Phi_{\bar{R} \bar{Q}}} = \bra{\Phi_{QR}} \Vert \Phi \rAngle$}.

Let $\rho_0\,{=}\,|\textrm{TFD} \rangle \langle \textrm{TFD} |$, the density matrix for the maximally entangled SYK code states in \eqnref{eq:H_TFD}. Under the Choi isomorphism, $\Vert \rho \rAngle\,{=}\,Z_\beta^{-1} e^{-\frac{\beta}{2} (H_R + H_{\bar R})} \Vert \Psi \rAngle$, where $H_R$ and $H_{\bar{R}}$ are SYK Hamiltonians acting on $\cH_R$ and $\bar{\cH}_R$, respectively.
Similarly, the noise channel $\cN$ can be mapped as $\hat{\cN}\,{=}\,C_\cN e^{-\cS_\cN}$, where $C_\cN$ is a normalization factor and $\cS_\cN$, 
for the noise channels in Eq.\eqref{eq:cN}, are given as
\begin{subequations}
\label{eq:Se}
\begin{align}
& \cS_{\cN^\hop_p} = -\phi_p \sum_{j} 2\im \gamma_j \bar{\gamma}_j\,,
\label{eq:Se_hop} \\
& \cS_{\cN^\inte_\interp} = - \phi_{\frac{\interp}{2N}} \Big(\sum_j 2\im \gamma_j \bar{\gamma}_j \Big)^2,
\label{eq:Se_int}
\end{align}
\end{subequations}
where $\phi_x\,{:=}\,\tanh^{-1} \frac{x}{1-x}$~\footnote{For Lindbladian evolution with jump operators being error operators, $\phi_x$ would correspond to the Lindbladian evolution time $t$.}.
Here $\gamma_i$ and $\bar{\gamma}_i$ are Majorana fermions operating on system $Q$ and $\bar{Q}$ respectively. With these operators, we get the decohered state
\begin{align}\label{eq:Erho}
     \Vert \cN [\rho] \rAngle  &= C_\cN e^{-\cS_\cN} \Vert \rho \rAngle  = C_\cN U_{\cN,\beta}  \Vert \Psi \rAngle  \,, \\
 U_{\cN,\beta} &:= e^{-\cS_\cN} e^{-\beta (H_R + H_{\bar R})/2} \,.
\end{align}

\begin{figure}[!t]
    \centering
    \includegraphics[width = 0.9\columnwidth]{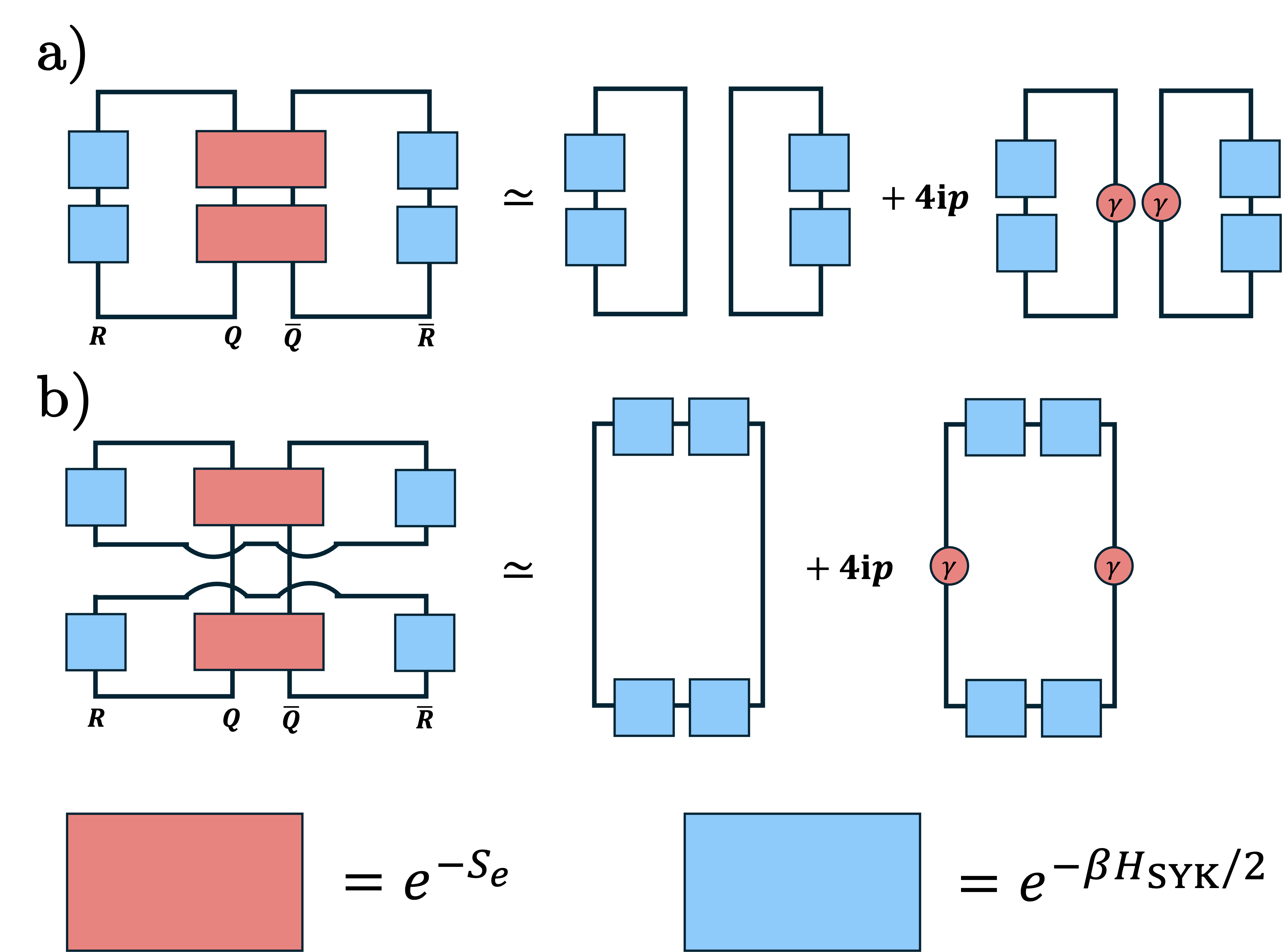}
    \caption{ {\bf Imaginary path integral diagram.} (a) $S_{QR}$ in Eq.\eqref{eq:SQR2} and (b) $S_Q$ in Eq.\eqref{eq:SQ2}. Each blue rectangle denotes an imaginary time evolution by the SYK Hamiltonian $e^{-\beta H_{\SYK}/2}$, and the orange rectangle denotes an evolution by the error channel. For weak hopping noise, the evolution by the error channel amounts to an insertion of the majorana bilinear $\sum_j\gamma_j \bar{\gamma}_j$. On the right side, we show their perturbative expansion to the first order for weak hopping noise. }
    \label{fig:Scircuit}
\end{figure}

We start with the evaluation of Renyi-2 entropies, which is obtained by contracting the doubled state $\Vert \cN[\rho] \rAngle$ with itself. 
Because the normalization factor $C_\cN$ appears as an additive term in both $S^{(2)}_{Q'}$ and $S^{(2)}_{Q'R}$ it will drop out of the expression for the coherent information. Thus, defining $\tilde{S}^{(2)}_a = S^{(2)}_a + 2\log C_\cN$, we have  $I_c^{(2)}\,{=}\,\tilde{S}^{(2)}_{Q'} \,{-}\, \tilde{S}^{(2)}_{Q'R}$ and 
\begin{subequations}
\label{eq:tildeSQ&QR}
\begin{align}
    \tilde{S}_{Q'R}^{(2)} &= -\log \lAngle \Psi \Vert U_{\cN,\beta}^2 \Vert \Psi \rAngle \label{eq:SQR2} \\
    \tilde S_{Q'}^{(2)} &= - \log \lAngle \Psi \Vert U_{\cN,\beta} \ket{\Phi_{R \bar{R}}}\hspace{-2pt}\bra{\Phi_{R\bar{R}}} U_{\cN,\beta} \Vert \Psi \rAngle. \label{eq:SQ2}
\end{align}
\end{subequations}
In Eq.\eqref{eq:SQ2} we inserted $\ket{\Phi_{R \bar{R}}} \hspace{-2pt} \bra{\Phi_{R \bar{R}}}$ to perform a partial trace over the subsystem $R$. 
The path integral that evaluates the above expression is represented diagrammatically in Fig.\ref{fig:Scircuit}.

In the large $N$ limit, these integrals are dominated by their saddle point solution and can be exactly computed for any finite temperature by numerically finding a self-consistent solution to the saddle point Schwinger-Dyson equations (see SI \ref{app:details}).
However, in practice, the convergence time increases rapidly with $\beta$, limiting our ability to access very low temperatures.
Consequently, our results are derived by extrapolating data from a range of finite temperatures to zero temperature.

\emph{Parity breaking noise.}---
We employ a perturbative approach for  $\cN^{\hop}_{p}$ with small $p$ to show that there is no error threshold for this decoherence channel. 
At zeroth order in $p$, $e^{-\tilde S_{Q'R}} \,{\simeq}\, Z_{\beta}^2$ and $e^{-\tilde S_{Q'}} \,{\simeq}\, Z_{2\beta}$, where $Z_\beta\,{=}\,\Tr e^{-\beta H_R}$. 
This is because the path integral of $\tilde S_{Q'R}$ has two separate contours of imaginary time evolution by $H_{\SYK}$ for a time $\beta$, while that of $\tilde S_{Q'}$ has a single contour for time $2\beta$, as illustrated in Fig.\ref{fig:Scircuit},

The correction to $\tilde S_{Q'R}$ vanishes at first-order in $p$. As seen in the rightmost circuit diagram in Fig.\ref{fig:Scircuit}(a), this correction factorizes into two disconnected contours with a single Majorana insertion in each contour. Each of the factorized contours then vanishes due to fermion parity symmetry.
On the other hand, $\tilde{S}_{Q'}$ has a nonzero first-order correction. As shown in Fig.\ref{fig:Scircuit}(b), this corresponds to the insertion of two Majorana fermions separated by the imaginary time $\beta$, which is just the Green's function $G_{2\beta}(\beta) = Z_{2\beta}^{-1}\braket{e^{-2\beta H_{\SYK}} \gamma(\beta) \gamma(0)}$. Putting this together, we have
\begin{subequations}
\begin{align}
    & \tilde S_{Q'R} (\cN^\hop_p) \simeq -2 \log Z_\beta - N \Gamma_{QR} \ p^2 \,,
    \label{eq:tildeSQR_hop} \\
    & \tilde S_{Q'} (\cN^\hop_p) \simeq - \log Z_{2\beta} - 4 N G_{2\beta}(\beta) p - N \Gamma_Q p^2 \,,
    \label{eq:tildeSQ_hop}
\end{align}
\end{subequations}
where $\Gamma_{QR,Q}$ are coefficients which we numerically find to be $\BigO(1)$ constants~\footnote{We remark that it is hard to find $\Gamma_{QR,Q}$ analytically, because it requires the computation of the shifted saddle point due to the evolution by the error channel of Eq.\eqref{eq:Se}}.

Now using the low-temperature scaling of the Green's function $G_{2\beta}(\beta)\sim \beta^{-2\Delta}$ and taking the temperature to be $T/J\sim N^{-\alpha}$, we obtain the  coherent information for weak parity breaking noise
\begin{equation}
    \frac{I_c^{(2)}(\cN^{\hop}_{p})}{N} \simeq s_0 - A_\Delta N^{-2\alpha\Delta} p - \big(\Gamma_Q - \Gamma_{QR} \big) p^2\,.
    \label{eq:Ic_hop}
\end{equation}
From this expression, we can directly solve for the $\epsilon$-threshold as defined above. For $\epsilon \sim N^{-\eta}$ it is 
\begin{align}
    p_{\thr}(\epsilon:\cN_{\hop}) \simeq \min \left\{N^{2\alpha\Delta-\eta}, \frac{N^{-\eta/2}}{(\Gamma_Q - \Gamma_{QR})^{1/2}} \right\} \,.
    \label{eq:pthr_hop}
\end{align}
$\Gamma_Q\,{-}\,\Gamma_{QR}$ is found numerically to be a positive constant of $\BigO(1)$. Therefore $p_{\thr}$ is upper-bounded by $N^{-\eta/2}$, which vanishes in the thermodynamic limit for any value of $\eta\,{>}\,0$. Taking $\eta\,{\rightarrow}\,0^+$, we lose an $\BigO(N)$ amount of coherent information at any finite error rate. Accordingly, there is no robust error threshold. This is indeed seen in Fig.\ref{fig:Main}(c) where we plot the coherent information as a function of the hopping error rate, obtained by numerically solving the saddle point equations to evaluate Eq.\eqref{eq:tildeSQ&QR}.

\emph{Parity conserving noise and spontaneous symmetry breaking.}---
We now show that there is a sharp error threshold for parity conserving decoherence.
To compute the Renyi entropies $S_{Q'}$ and $S_{Q'R}$ with noise level $q$, we perform a Hubbard-Stratonovich decoupling of the quartic term in the action \eqnref{eq:Se_int} of the quantum channel
\begin{equation}
   e^{-\cS_{\cN^\inte_\interp}} = \int d\phi \, e^{-\frac{N}{2\interp} \phi^2 - \phi \sum_j 2\im \gamma_Q^j \gamma_{\overline{Q}}^j}.
\end{equation}
where we used that $\phi_{q/2N}\,{\approx}\,q/2N$ for $N\,{\rightarrow}\,\infty$.
In the RHS, the second term in the exponent is the action of the parity breaking channel with strength set by the Hubbard-Stratonovich field $\phi$. 
Now we can integrate out the fermions and take the large $N$ saddle point by minimizing with respect to $\phi$ to get the Renyi entropies 
\begin{equation}
\begin{split}
    & \tilde S^{(n)}_{A} (\cN^\inte_{\interp}) := \textrm{min}_{\phi} \left\{\frac{n N}{2\interp} \phi^2 + \tilde S^{(n)}_{A} \Big(\cN^\hop_{p_\phi} \Big) \right\} \,
    \label{eq:Sint}
\end{split}
\end{equation}
for $A\,{\in}\,\{Q',Q'R\}$, and $p_\phi\,{:=}\,\tanh \phi/(1+ \tanh \phi)$. It now becomes clear how an error threshold emerges in the problem. For sufficiently small $q$ at $T\,{=}\,0$, we expect the symmetric saddle point $\phi=0$, which would leave the coherent information as in the pure system. Above a critical value of $q\,{>}\,q_c$, $\phi$ may develop an expectation value $\langle\phi\rangle\,{\ne}\,0$ in the saddle point of $S_Q(\cN_q^\inte)$. This amounts to spontaneous breaking of strong symmetry down to weak symmetry~\cite{Lee2023PRXQ, StW_U(1), sala2024spontaneousstrongsymmetrybreaking, lessa2024strongtoweakspontaneoussymmetrybreaking}. The spontaneous breaking of strong fermion parity symmetry gives rise to an effective parity breaking decoherence term that continuously degrades the coherent information. 

To study the threshold explicitly, we substitute the perturbative solutions Eqs.\eqref{eq:tildeSQR_hop} and \eqref{eq:tildeSQ_hop} for $\tilde S_A(\cN^\hop_{p(\phi)} )$ in \eqref{eq:Sint}. Minimizing with respect to $\phi$ we get
\begin{equation}
\begin{split}
    & \tilde S_{Q'R}\big(\cN^\inte_{\interp}\big) = -2\log Z_{\beta} \\
    & \tilde S_{Q'}\big(\cN^\inte_{\interp}\big) = -\log Z_{2\beta} -\frac{ 4N G_{2\beta}(\beta)^2 }{\interp^{-1} - \Gamma_Q}
    \label{eq:tildeSint_lowT}
\end{split}
\end{equation}
Substituting this into \eqref{eq:tildeSint_lowT} gives 
\begin{equation}
    \frac{I_{c}^{(2)}(\cN_{\interp}^\inte)}{N} \simeq s_0 - \frac{N^{-4\alpha \Delta} }{\interp^{-1} - \Gamma_Q}.
    \label{eq:Ic_int}
\end{equation}
We see that in the thermodynamic limit, the quantum information is protected as long as $q\,{<}\,1/\Gamma_Q$. More precisely we can solve for the $\epsilon$-threshold with $\epsilon \sim N^{-\eta}$ to get
\begin{align}
    p_{\thr}(\epsilon:\cN_{\inte}) \simeq \frac{1}{\Gamma_Q + N^{\eta-4\alpha\Delta}}\,.
    \label{eq:pthr_int}
\end{align}
This expression implies a non-vanishing threshold in the thermodynamic limit for $\eta\,{<}\,4\alpha\Delta$. If we scale $T$ to zero as $1/N$, i.e. $\alpha=1^-$, then we can ensure an $\epsilon$ threshold for $\epsilon> 1/N^{4\Delta}$.

In the low-rank SYK models, $\Delta$ can be tuned continuously in the range $\Delta\,{\in}\,(1/4,1/2)$. Therefore in this system, we can have an $\epsilon$-threshold for $\epsilon\,{\sim}\,1/N$, namely a threshold against losing a single qubit. The standard 
SYK model with $\Delta\,{=}\,1/4$ is the marginal case for such a threshold, but we can certainly protect against any subextensive $\cO(N^{1-\eta>0})$ loss of coherent information. 

So far we discussed perturbative results in the limit of weak decoherence, which does not allow us to compute how the coherent information decreases above the threshold. To this end, we carry out an exact numerical solution of the saddle equation \eqnref{eq:Sint}. The results shown in Fig.\ref{fig:Main}(d) confirm the existence of a sharp error threshold in both the low-rank and standard SYK models and show a continuous degradation of the coherent information beyond this point.

\emph{Replica Limit.}---
The calculation of Renyi-2 coherent information can be straightforwardly extended to higher-Renyi quantities as elaborated in the Supplementary material, see \appref{sec:Renyi3}.
Again, the existence of a robust error threshold can be demonstrated following a perturbative approach as before. For parity breaking noise $\cN_{p}^{\hop}$, to leading order in $p$, we find, 
\begin{align}
    \tilde S_{Q'R}^{(n)}(\cN_{p}^\hop) & \simeq - n \log Z_\beta - N \Gamma_{QR}^{(n)}p^2 \,, \\
    \tilde S_{Q'}^{(n)}(\cN_{p}^\hop) & \simeq - \log Z_{n \beta} - 2 n N G_{n\beta}(\beta) - N \Gamma_{Q}^{(n)}p^2 \,,
    \label{eq:Renyi_n}
\end{align}
Using these expressions, the coherent information under parity conserving noise can be derived as in \eqnref{eq:Sint}, 
which gives $I_c^{(n)}/N \simeq s_0 - N^{-4 \alpha \Delta}/( n (2q)^{-1} - \Gamma_Q^{(n)})$.
As long as $\Gamma_Q^{(n)}$ is an $\BigO(1)$ constant, there is a robust error threshold against parity conserving noise.
We confirm numerically that this is indeed the case for $n=3$. Moreover, we find that the threshold for $n=3$ is larger than for $n=2$, similar to the findings in the case of topological codes~\cite{fan2023diagnostics, Lee2023PRXQ, bao2023mixedstate, lyons2024understanding, su2024tapestry}. 
We expect that the extrapolation $n\to 1$ yields an error threshold also for the von Neumann coherent information, which carries information theoretic significance.

\emph{Discussion.}---In this paper, we investigated the decoherence threshold in a family of SYK models that can serve as approximate quantum codes with the number of logical qubits $k\propto N$, encoded into the extensive quasi-ground state manifold.   
Using both numerical and analytical studies, we found that low-rank SYK codes are robust against parity-conserving noise but not against parity-breaking noise.
The threshold behavior of the former can be understood as a phase transition involving spontaneous symmetry breaking~\cite{Lee2023PRXQ, StW_U(1), sala2024spontaneousstrongsymmetrybreaking, lessa2024strongtoweakspontaneoussymmetrybreaking} of the fermion parity from strong to weak symmetry.

An interesting question raised by our results is to what extent they can be extended to qLDPC codes. While SYK-type approximate codes resemble good qLDPC codes in their ability to encode many logical qubits and exhibit some robustness, they lack sparsity. Exploring sparse SYK models~\cite{xu2020sparsemodelquantumholography} could provide more direct insights into the nature of good qLDPC codes~\cite{panteleev2022asymptotically, dinur_qldpc_decoder}.

Another intriguing question concerns the interpretation of the error threshold and the transition from strong to weak spontaneous symmetry breaking (SSB) in the dual gravity picture of the SYK model and its connection to traversable wormholes~\cite{Maldacena_2017, gao2021traversablewormholeteleportationprotocol}.
Such an investigation could deepen our understanding of the interplay between quantum error correction and holographic duality.
Finally, we note that transition behaviors has been observed in SYK models coupled to the environment, involving path integrals with non-Hermitian operators~\cite{Garc_a_Garc_a_2022, PhysRevD.107.066007, PhysRevB.108.075110}. This contrasts with our approach, which involves purely imaginary path integrals with Hermitian operators. Elucidating the connection between the two distinct approaches is an interesting direction for future work.

\acknowledgments
We thank Gregory Bentson, Soonwon Choi, Thomas Faulkner, Ruihua Fan, Samuel Garratt, Luca Illisieu, Zi-Wen Liu, Lucas Sa, Brian Swingle, and Zack Weinstein for fruitful discussions and comments. 
The work is supported by the Simons Investigator Award and a faculty startup grant at the University of Illinois, Urbana-Champaign.

\bibliography{ref}

\mbox{~}
\clearpage
\newpage

\onecolumngrid

\makeatletter
\def\l@subsection#1#2{}
\def\l@subsubsection#1#2{}
\makeatother

\setcounter{equation}{0}
\setcounter{figure}{0}
\setcounter{table}{0}

\makeatletter
\renewcommand{\theequation}{S\arabic{equation}}
\renewcommand{\thefigure}{S\arabic{figure}}
\setcounter{subsection}{0}

\begin{center}
    \textbf{\large Supplementary Material for \\ \vspace{7pt}
    ``Error Threshold of SYK Codes from Strong-to-Weak Parity Symmetry Breaking''}
    
    \vskip6mm
    
    \end{center}

\vspace{10pt}
    
In this supplementary material, we elaborate on several technical details that support the claims made in the manuscript.

\tableofcontents 

\section{Crash Course into SYK type Models}
\label{app:SYK}
In this section, we provide a brief introduction into the SYK and low-rank SYK model.
As explained in the main text, both models are essentially composed of all to all four fermion interactions as given by the following Hamiltonian,
\begin{equation*}
    H = \sum_{ijkl} J_{ijkl} \chi_i \chi_j \chi_k \chi_l
\end{equation*}
In the SYK model, $J_{ijkl}$ is a totally random anti-symmetric tensor of zero mean and variance $\J^2/6N^3$.
Correspondingly, if we were to decompose this tensor into two parts by grouping $ij$ together and $kl$ together, the resulting matrix;s rank would be $N^2$.
In contrast, the low-rank SYK model's random tensor $J$ has some more structure, and is expressed as,
\begin{equation*}
    J_{ij,kl} = \sum_{i=1}^{\gam N} u_{ij}^n u_{kl}^n
\end{equation*}
where $u^n$ are real random matrices of zero mean and variance $g^2/N^2$.
As explicitly stated above in the construction, the low-rank SYK model's random matrix $J$ is of rank $\gam N$.

Both models become conformal at low temperatures $\beta J \gg 1$, and the solution to the saddle-point equations give rise to a fermion Green's function $G_\beta(\tau) = \braket{e^{-\beta H}\chi(\tau) \chi(0)}$ of the following form:
\begin{equation}
    G_\beta(\tau) = A_\Delta \left(\frac{\pi}{\beta \sin{\frac{\pi\tau}{\beta}}}\right)^{2\Delta}
    \label{eq:appGSYK}
\end{equation}
Fermions in the SYK model obtains a scaling dimension $\Delta = 1/4$.
On the other hand, those in the low-rank SYK model obtains a scaling dimension $\Delta \in [0,1/2]$, which depends on the rank as \cite{lowrank},
\begin{equation}
\gamma = \frac{(2\Delta-1)(\sec 2\pi \Delta - 1)}{8\Delta - 2} \,.
\label{eq:gammadelta}
\end{equation}
In the sub-extensive limit of $\gam \rightarrow 0$, $\Delta \rightarrow \frac12$.
On the other hand, in the super-extensive limit of $\gam \gg 1$, $\Delta \rightarrow \frac14$ and we arrive at the SYK model.

Another feature of these SYK-type models is that they have an extensive zero temperature entropy \cite{lowrank}.
$s_0$ is a monotonically increasing function of $\gam$, and in the limit of $\gam \rightarrow \infty$, $s_0$ saturates to that of the SYK model, 0.233.

\section{Doubled State Path Integral Formalism}
\label{app:details}
In this section, we derive the doubled state path integrals approach that we utilized to find the Renyi-2 entropies $S_{Q'R,Q'}^{(2)}$, and find the asymptotic behavior of the Renyi-2 entropies at low temperatures.

\subsection{Parity Breaking Noise}
We first derive the doubled state path integrals for parity breaking noise.
Let us first derive $\tilde{S}^{(2)}_{Q'R}$.
Using Eq.\eqref{eq:SQR2}, it is given by,
\begin{equation}
\begin{split}
    \exp\Big\{ -\tilde{S}^{(2)}_{Q'R}(\cN_p^\hop) \Big\} &= \lAngle \Psi \Vert  e^{2\phi_p \sum_i 2\im \gamma_Q^i \gamma_{\bar Q}^i} e^{-\beta (H_R + H_{\bar R})} \Vert \Psi \rAngle \textrm{ where } \tanh \phi_p = \frac{p}{1-p} \,, \\
    &= \lAngle \Psi \Vert  e^{\beta\mu \sum_i 2\im \gamma_Q^i \gamma_{\bar Q}^i} e^{-\beta (H_R + H_{\bar R})} \Vert \Psi \rAngle \textrm{ where } \beta\mu = 2\phi_p\,.
    \label{eq:appSQR0}
\end{split}
\end{equation}
To evaluate $\tilde{S}^{(2)}_{Q'R}(\cN_p^\hop)$, we can perform a coherent state path integral, and insert grassmanian fields $\gamma_{R,\bar R}(\tau), \gamma_{Q,\bar Q}(\tau)$ with $\tau$ ranging between 0 and $\beta$, with $\tau = 0$ denoting the coherent states at $\Vert \Psi \rAngle$.
At the boundaries at $\tau = \beta$ where the grassmanian fields meet $\lAngle \Psi \Vert$ or $\Vert \Psi \rAngle$, we must have
$\gamma_{R,\bar R}(\beta) = \mp \im \gamma_{Q,\bar Q}(\beta)$.
Instead of keeping track of four fermion fields, however, we stitch together $\gamma_{R,\bar R}(\tau)$ with $\gamma_{Q, \bar Q}(\tau)$ at the boundaries of $\tau = 0, \beta$ and define a new two component fermion field $\bm \psi = (\psi_1, \psi_2)$, where $\psi_1$ stands for the grassmanian fields living on the ket side contour ($Q$ and $R$), and $\psi_2$, those on the bra side contour ($\bar Q$ and $\bar R$).
They are each given in terms of the original coherent states $\gamma_{R,\bar R, Q, \bar Q}$ as,
\begin{equation}
\begin{split}
    & \psi_1(\tau) = \begin{cases}
    \gamma_R(\tau) \textrm{ for } 0 \le \tau < \beta \\
    -\im\gamma_Q(2\beta - \tau) \textrm{ for } \beta \le \tau < 2\beta
    \end{cases} \\
    & \psi_2(\tau) = \begin{cases}
    \gamma_{\bar R}(\tau) \textrm{ for } 0 \le \tau < \beta \\
    \im\gamma_{\bar Q}(2\beta - \tau) \textrm{ for } \beta \le \tau < 2\beta
    \end{cases}
    \label{eq:apppsi12}
\end{split}
\end{equation}
In turn, Eq.\eqref{eq:appSQR0} can be expressed as,
\begin{equation}
\begin{split}
    & \tilde{S}^{(2)}_{Q'R}(\cN_p^\hop) = -\log \int [\cD \bm{\psi} ] e^{- \bcS_{QR}[ \bm{\psi} ]} \textrm{ where } \\
    & \bcS_{QR}[ \bm{\psi} ] = \sum_\sigma \int_0^{2\beta} \hspace{-2pt} d\tau \, (\psi_\sigma  \partial_\tau \psi_\sigma  + g_\tau H_\sigma + \tilde g_\tau H_\textrm{err} )  \\
    & \ \ H_\sigma = \sum_{ijkl} J_{ijkl} \psi_\sigma^i \psi_\sigma^j \psi_\sigma^k \psi_\sigma^l, \quad H_\err = 2 \im \mu \sum_j \psi_1^j \psi_2^j
    \label{eq:appSQR}
\end{split}
\end{equation}
where $g_\tau = \Theta(\beta-\tau)$ and $\tilde{g}_\tau = 1- g_\tau$ enforces the proper imaginary time evolution by the SYK-type / error Hamiltonian, respectively.
To evaluate \eqref{eq:appSQR}, we utilize the power of large $N$.
We first define the Green's function $G_{\sigma\sigma'}(\tau,\tau') = \braket{\psi_\sigma(\tau)\psi_{\sigma'}(\tau')}$ and enforce it by inserting a dirac delta function in the following manner.
\begin{equation}
\begin{split}
    \delta\Big(\sum_{i} \psi^i_{\sigma}(\tau) \psi^i_{\sigma'}(\tau') - N G_{\sigma \sigma'}(\tau,\tau')\Big) =& \int [\cD\bm\Sigma] \exp\Bigg(\Sigma_{\sigma \sigma'}(\tau,\tau')\Big(\sum_{i} \psi^i_{\sigma}(\tau) \psi^i_{\sigma'}(\tau') - N G_{\sigma \sigma'}(\tau, \tau')\Big)\Bigg) \,.
\end{split}
\end{equation}
After disorder averaging, and integrating out the fermions, the effective action then becomes,
\begin{equation}
\begin{split}
    \bcS_{QR}(G,\Sigma)/N &= -\frac12 \log \det \Big\{ \delta_{\sigma \sigma'} \delta(\tau-\tau') \partial_{\tau'} - \Sigma_{\sigma \sigma'}(\tau,\tau') - 2\mu \tilde g_\tau \textbf{Y}_{\sigma \sigma'} \delta(\tau-\tau') \Big\} \\
    & \quad \quad + \frac12 \int_{\tau, \tau'} \Sigma_{\sigma \sigma'}(\tau, \tau') G_{\sigma \sigma'}(\tau, \tau') - \frac{\J^2}{2q} g_\tau g_{\tau'} G_{\sigma \sigma'}^4(\tau, \tau') \,.
    \label{eq:appSQReff}
\end{split}
\end{equation}
Here, $\textbf{Y}$ denotes the Pauli Y matrix.
This action is dominated by the saddle-point solution due to the large $N$ scaling.
Taking a functional derivative of Eq.\eqref{eq:appSQReff}, the corresponding saddle-point equations are given as,
\begin{equation}
\begin{split}
    & G_{\sigma \sigma'}(\tau, \tau') = \Big\{ \delta_{\sigma \sigma'} \delta(\tau-\tau') \partial_{\tau'} - \Sigma_{\sigma \sigma'}(\tau,\tau') - 2\mu \tilde g_\tau \textbf{Y}_{\sigma \sigma'} \delta(\tau-\tau') \Big\}^{-1}_{\sigma \sigma', \tau \tau'} \,, \\
    & \Sigma_{\sigma \sigma'}(\tau, \tau') = g_\tau g_{\tau'} J^2 G^3_{\sigma \sigma'}(\tau, \tau') \,.
    \label{eq:appSD_LR}
\end{split}
\end{equation}
Finally, \eqref{eq:appSD_LR} can be solved numerically to obtain the fermion Green's function $G_{\sigma\sigma'}$.
Subsequently, the solution that we obtain can be inserted back into \eqref{eq:appSQReff} to obtain the value of the path integral $\tilde S_{Q'R}$.

We now turn to $S_{Q'}^{(2)}$.
Similar to before,
\begin{equation}
    \exp\Big\{ -\tilde{S}_{Q'}^{(2)}(\cN^\hop_p) \Big\} = \lAngle \Psi \Vert e^{-\beta (H_R + H_{\bar R})/2} \ket{\Phi_{R\bar R}} e^{\beta \mu \sum_i 2\im \gamma_Q^i \gamma_{\bar Q}^i} \bra{\Phi_{R\bar R}} e^{-\beta (H_R + H_{\bar R})/2} \Vert \Psi \rAngle \,.
    \label{eq:appSQ0}
\end{equation}
As above, we now insert grassmanian fields $\gamma_{R,\bar R}(\tau), \gamma_{Q,\bar Q}(\tau)$ with $\tau = 0$ at $\ket{\Phi_R}$.
Again, instead of keeping track of four fermion fields, we stitch them altogether to form a single fermion field $\psi$.
With the boundary conditions at $\lAngle \Psi \Vert$, $\Vert \Psi \rAngle$, $\ket{\Phi_R}$ and $\ket{\Phi_R}$, the fermion field $\psi$ is related to the original grassmanian fields $\gamma_{R,\bar R}$ and $\gamma_{Q,\bar Q}$ by,
\begin{equation}
    \psi(\tau) =
    \begin{cases}
        \gamma_R(\tau) \textrm{ for } 0 \le \tau < \beta/2 \\
        -\im\gamma_{\bar R}(\beta - \tau) \textrm{ for } \beta/2 \le \tau < \beta \\
        \gamma_{\bar Q}(\tau - \beta) \textrm{ for } \beta \le \tau < 2 \beta \\
        \im \gamma_{\bar R}(5\beta/2-\tau) \textrm{ for } 2\beta \le \tau < 5 \beta/2 \\
        \gamma_R(\tau - 5\beta/2) \textrm{ for } 5\beta/2 \le \tau < 3 \beta \\
        \im \gamma_Q(4\beta - \tau) \textrm{ for } 3 \beta \le \tau < 4\beta \\
    \end{cases}
    \label{eq:apppsi}
\end{equation}
In turn Eq.\eqref{eq:appSQ0} translates to,
\begin{equation}
\begin{split}
    & \tilde{S}_{Q'}^{(2)}(\cN^\hop_p) = -\log \int D\psi \exp\{-\bcS_{Q}(\psi)\} \,, \\
    & \bcS_{Q}(\psi) = \int_0^{4\beta} d\tau \psi \partial_\tau \psi  + f_\tau H + \tilde f_\tau H_{err} \,, \\
    & H = \sum_{ijkl} J_{ijkl} \psi^i \psi^j \psi^k \psi^l \,, \ H_\err = -2 \mu \sum_j \psi^j(5\beta-\tau) \psi^j(\tau)
    \label{eq:appSQ}
\end{split}
\end{equation}
Where $f_\tau = \Theta(\beta-\tau) + \Theta(\tau-2\beta) \Theta(3\beta-\tau)$ and $\tilde f_\tau = \Theta(2\beta-\tau)\Theta(\tau-\beta)$ enforces the proper imaginary time evolution.

As before, upon introducing the fermion Green's function $G(\tau,\tau') = \braket{\psi(\tau)\psi(\tau')}$ with its corresponding self-energy $\Sigma$, disorder averaging, and integrating out the fermions, we find that $\bcS_Q$'s effective $G-\Sigma$ action.
It is given by,
\begin{equation}
\begin{split}
    \bcS_Q(G,\Sigma)/N &= -\frac12 \log \det \Big\{\delta(\tau-\tau') \partial_{\tau'} - \Sigma(\tau,\tau') + 2 \mu \tilde f_\tau \sgn(\tau-\tau') \delta(\tau+\tau'-5\beta) \Big\} \\
    & \quad + \frac12 \int_{\tau, \tau'} \Sigma(\tau, \tau') G(\tau, \tau') - \frac{J^2}{2q} f_\tau f_{\tau'} G^4(\tau, \tau') \,.
    \label{eq:appSQeff}
\end{split}
\end{equation}
The large $N$ saddle point of Eq.\eqref{eq:appSQeff} is given as,
\begin{equation}
\begin{split}
    & G(\tau, \tau') = \Big\{\delta(\tau-\tau') \partial_{\tau'} - 2 \mu \tilde f_\tau \sgn(\tau-\tau') \delta(\tau + \tau' - 5\beta) - \Sigma(\tau,\tau') \Big\}^{-1}_{\tau,\tau'} \,, \\
    & \Sigma(\tau, \tau') = f_\tau f_{\tau'} J^2 G^3(\tau, \tau') \,.
    \label{eq:appSD_L}
\end{split}
\end{equation}
Eq.\eqref{eq:appSD_L} can be evaluated numerically to obtain the Green's function, and inserted back into \eqref{eq:appSQeff} to obtain the value of the path integral $\tilde S_{Q'}$.

\subsection{Parity Conserving Noise}
We now turn to parity conserving noise.
Our main object will be to derive \eqref{eq:Sint}, which allowed us to find the entanglement entropies after undergoing parity conserving decoherence from the entanglement entropies after undergoing parity breaking decoherence.
Let us again begin with the deriving $\tilde S_{Q'R}^{(2)}$. From \eqref{eq:SQR2}, it is given by,
\begin{equation}
\begin{split}
    \exp\Big\{ - S_{QR}^{(2)} \Big\} &= C_\cN^2 \bra{\rho_\TFD} \prod_{ij} e^{-8\interp/N \gamma_Q^i \gamma_Q^j \gamma_{\bar Q}^j \gamma_{\bar Q}^i} \ket{\rho_\TFD} \\
    &= C_\cN^2 \lAngle \Psi \Vert e^{\interp/N \big(\sum_i 2\im \gamma_Q^i \gamma_{\bar Q}^i\big)^2} e^{-\beta (H_R + H_{\bar R})} \Vert \Psi \rAngle \\
    &= C_\cN^2 \sqrt{N/\pi\interp} \int_\phi e^{-\frac{N}{\interp} \phi^2} \lAngle \Psi \Vert e^{\phi \sum_{i} 2\im \gamma_Q^i \gamma_{\bar Q}^i} e^{-\beta (H_R + H_{\bar R})} \Vert \Psi \rAngle \,.
    \label{eq:appSQRinte00}
\end{split}
\end{equation}
where in the third line we have done a Hubbard-Stratonovich decomposition.
Consequently, after absorbing $\sqrt{N/\pi\interp}$ into the proportionality constant $C_\cN$, we find,
\begin{equation}
\begin{split}
    \tilde S_{Q'R}^{(2)} (\cN_{\interp}^\inte) &= -\log \int_\phi e^{-\frac{N}{\interp} \phi^2} \lAngle \Psi \Vert e^{\phi \sum_{i} 2\im \gamma_Q^i \gamma_{\bar Q}^i} e^{-\beta (H_R + H_{\bar R})} \Vert \Psi \rAngle \\
    &= -\log \int_\phi \exp{-\frac{N}{\interp} \phi^2-\tilde S^{\hop}_{QR}\big(p(\phi)\big)} \,, \textrm{ where } \phi = \tan^{-1}\frac{p(\phi)}{1-p(\phi)} \,.
    \label{eq:appSQRinte0}
\end{split}
\end{equation}
Due to the large $N$ limit, $\tilde S_{Q'R}^{(2)} (\cN_{\interp}^\inte)$ becomes the minimum value of the exponent in Eq.\eqref{eq:appSQRinte0}, and we find,
\begin{equation}
\tilde S_{Q'R}^{(2)}(\cN^\inte_{\interp}) := {\min}_\phi \Bigg\{ \frac{1}{\interp} \phi^2 + \tilde S_{Q'R}^{(2)} (\cN^\hop_{p(\phi)}) \Bigg\} \,.
\label{eq:appSQRinte1}
\end{equation}
It is not too difficult to show the same for $S_{Q'}^{(2)}$.

\subsection{Both Noise}
We now turn to the case in which we have both types of decoherence channels, the parity breaking noise with an error rate $p$, and the parity conserving noise with an error rate $\interp/N$.
Let us again first find $S_{Q'R}^{(2)}$.
From the two earlier subsections of this section, we have,
\begin{equation}
\begin{split}
\exp{-S_{Q'R}^{(2)}} &= C_\cN^2 \bra{\rho_{\TFD}} \prod_i e^{\phi_0 2\im \gamma_Q^i \gamma_{\bar Q}^i} \prod_{ij} e^{-8\interp/N \gamma_Q^i \gamma_Q^j \gamma_{\bar Q}^j \gamma_{\bar Q}^i} \ket{\rho_{\TFD}} \textrm{ where } \phi_0 = \tanh^{-1} \frac{p}{1-p} \\
&= C_\cN^2 \int_\phi e^{-\frac{N}{\interp} \phi^2} \lAngle \Psi \Vert e^{(\phi + \phi_0) \sum_{i} 2\im \gamma_Q^i \gamma_{\bar Q}^i} e^{-\beta (H_R + H_{\bar R})} \Vert \Psi \rAngle \,.
\end{split}
\end{equation}
Consequently, as in Eq.\eqref{eq:appSQRinte1}, we find that
\begin{align}
    & \tilde S_{Q'R,Q'}^{(2)} (\cN^{\both}_{p_\hop, \interp}) = \min \left\{\frac{1}{\interp} \phi^2 + \tilde S^\hop_{QR,Q} (\cN^{\hop}_{p_{(\phi+\phi_0)}}) \right\} \,, \\
    & \textrm{ where } \phi_0 = \tanh^{-1} \frac{p_\hop}{1-p_\hop} \,.
    \label{eq:appSboth}
\end{align}
One can show the same for $\tilde S_{Q'}^{(2)}$.

\begin{figure}
    \centering
    \includegraphics[width=0.75\linewidth]{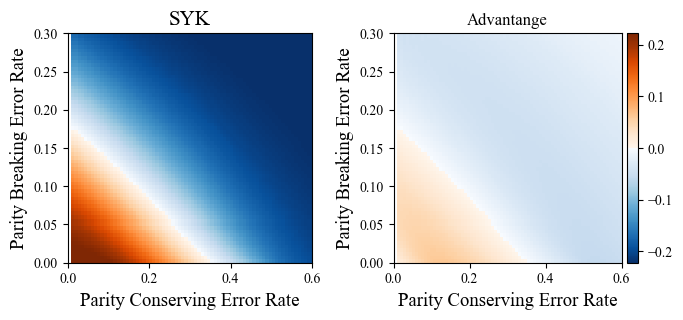}
    \includegraphics[width=0.75\linewidth]{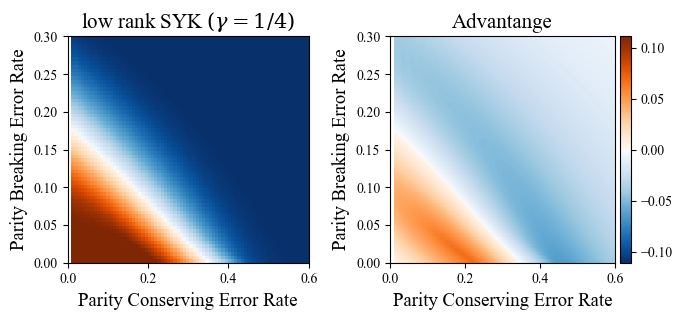}
    \caption{Color plot of the Renyi-2 coherent information and relative advantage against no encoding as a function of parity breaking / conserving noise for the \textit{(Top)} SYK model, and \textit{(Bottom)} low-rank SYK model, $\gam = 1/4$.}
    \label{fig:CohInfo_Renyi2both}
\end{figure}

Taking their difference, we find the Renyi-2 coherent information against both types of noise, which is plotted in the left column figures of Fig.\ref{fig:CohInfo_Renyi2both} for the SYK model and for the low-rank SYK model at $\gam = 1/4$.
The coherent information decreases continuously as a function of both noises.
However, there is a clear advantage over no encoding, especially so for the low-rank SYK model.
To highlight this, we plot on the right column of Fig.\ref{fig:CohInfo_Renyi2both} the difference between the coherent information for the SYK-type models, and that of Bell pairs without any encoding.

\section{Renyi-3 Coherent Information} \label{sec:Renyi3}
In this section, we explain how to obtain the Renyi-3 coherent information through a extension of the formalism that we have detailed in the main text and in SI.\ref{app:details}.
As we shall demonstrate, we see qualitatively the same behaviors in the Renyi-3 coherent information as that with the Renyi-2 calculations, such as a continuous decrease of coherent information under parity breaking noise, and a robust protection of coherent information under parity conserving noise.

Since the Renyi-3 entanglement entropy is defined as, $S^{(3)} = -\frac12 \log \tr \rho^3$, we need to find $\tr \rho_{Q,QR}^3$.
To this end, we first find the expression for $\tr \rho^3$ in the doubled state representation.
This step is necessary since the decohered density matrix has a simple form in the doubled state representation of Eq.\eqref{eq:Erho}.

Let us first prepare three copies of the density matrices in the doubled state representation: $\Vert \rho \rAngle_{a}$, with $a$ ranging from 1 to 3.
It is easy to see that,
\begin{equation}
    \tr \rho^3 = \lAngle \Phi_{\bar{3}1} \Vert \lAngle \Phi_{3\bar{2}} \Vert \lAngle \Phi_{2\bar{1}} \Vert \Vert \rho \rAngle_{1} \Vert \rho \rAngle_{2} \Vert \rho \rAngle_{3}
    \label{eq:trrho3}
\end{equation}
where $\lAngle \Phi_{a b} \Vert$ is defined as the maximally entangled state between the fermions of $a$ and the fermions of $b$ that satisfies,
\begin{equation}
\lAngle \Phi_{a b} \Vert (\gamma_a + \im \gamma_b) = 0 \,,
\end{equation}
and the presence (absence) of the bar denotes the bra (ket) fermions.
Contracting with $\lAngle \Phi_{2 \bar 1} \Vert$ for example amounts to identifying the bra of $\Vert \rho \rAngle_{1}$ with the ket of $\Vert \rho \rAngle_{2}$.
Ergo, repeating this contraction process results in $\tr \rho^3$.

\begin{figure}
    \centering
    \includegraphics[width=0.8\linewidth]{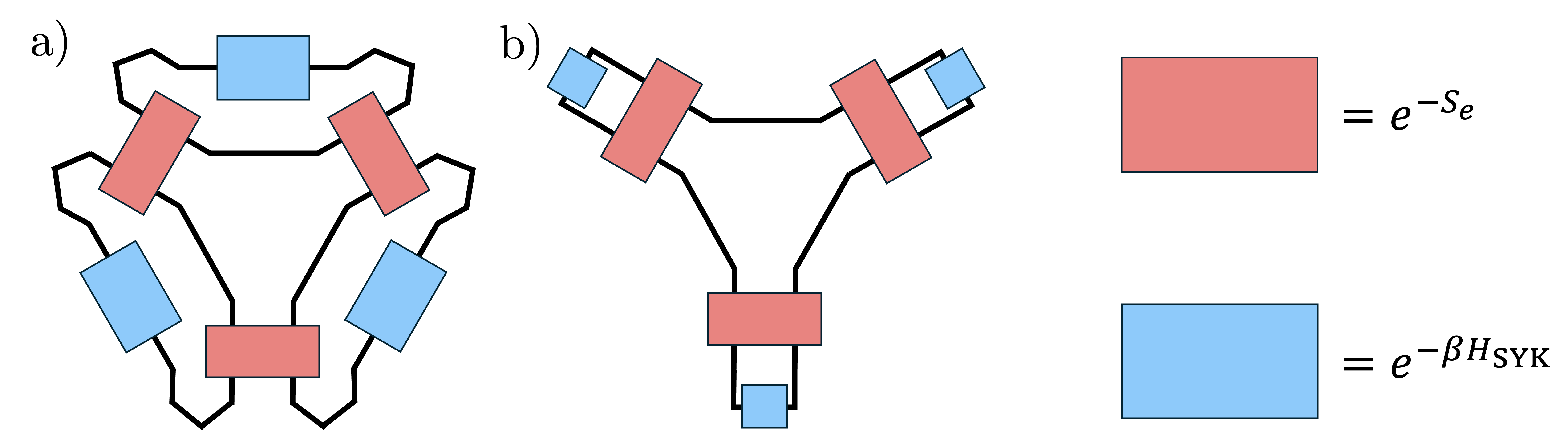}
    \caption{Circuit diagram of the path integrals of Eq.\eqref{eq:Renyi3SQR} and Eq.\eqref{eq:Renyi3SQ}. Blue rectangles denote evolution by the SYK Hamiltonian, $e^{-\beta H_{\SYK}}$, and the orange rectangles, that by the error Hamiltonian, $e^{-\cS_\cN}$.}
    \label{fig:Renyi3circuit}
\end{figure}

Just as with Renyi-2 entropies, $S_{Q',Q'R}^{(3)}$ can be expressed through a path integral formalism.
Whereas with Renyi-2, $S_{Q'R}^{(2)}$ had two contours, Renyi-3 $S_{Q'R}^{(3)}$ now has three contours as depicted in Fig.\ref{fig:Renyi3circuit}.
Defining $\tilde S_{a} = S_{a} + 3 \log C_\cN$ to simplify the contribution from the normalization factor,  $\tilde S_{Q'R}^{(3)}(\cN_p^\hop)$ is given by the following path integral of,
\begin{equation}
\begin{split}
    & \tilde{S}_{Q'R}^{(3)}(\cN_p^\hop) = -\log \int [\cD \bm{\psi} ] e^{- \bcS_{QR}[ \bm{\psi} ]} \textrm{ where } \\
    & \bcS_{QR}[ \bm{\psi} ] = \sum_{\sigma = 1}^3 \int_0^{2\beta} \hspace{-2pt} d\tau \, (\psi_\sigma  \partial_\tau \psi_\sigma  + g_\tau H_\sigma + \tilde g_\tau H_\textrm{err} )  \\
    & \ \ H_\sigma = \sum_{ijkl} J_{ijkl} \psi_\sigma^i \psi_\sigma^j \psi_\sigma^k \psi_\sigma^l, \\
    & \ \ H_\err = 
    2 \mu \sgn(3\beta-2\tau) \sum_j \psi_{3}^j(3\beta-\tau) \psi_{1}^j(\tau) + \psi_{2}^j(3\beta-\tau) \psi_{3}^j(\tau) + \psi_{1}^j(3\beta-\tau) \psi_{2}^j(\tau) \,.
    \label{eq:Renyi3SQR}
\end{split}
\end{equation}
Where $g_\tau = \Theta(\beta-\tau)$ and $\tilde g_\tau = \Theta(\tau-\beta)$ to ensure the proper imaginary time evolution.

Similarly, $\tilde S_{Q'}^{(3)}(\cN_p^\hop)$ is given as,
\begin{equation}
\begin{split}
    & \tilde{S}_{Q'}^{(3)}(\cN_p^\hop) = -\log \int [\cD \bm{\psi} ] e^{- \bcS_{QR}[ \bm{\psi} ]} \textrm{ where } \\
    & \bcS_{Q}[ \psi ] = \int_0^{6\beta} \hspace{-2pt} d\tau \, (\psi  \partial_\tau \psi  + f_\tau H_\sigma + \tilde f_\tau H_\textrm{err} )  \\
    & \ \ H_\sigma = \sum_{ijkl} J_{ijkl} \psi^i \psi^j \psi^k \psi^l, \\
    & \ \ H_\err(\tau) = \begin{cases}
    2 \mu \sgn(7\beta-2\tau) \sum_j \psi^j(7\beta-\tau) \psi^j(\tau) \textrm{ for } \beta \le \tau < 3\beta/2 \textrm{ and } 11\beta/2 \le \tau < 6\beta \\
    2 \mu \sgn(5\beta-2\tau) \sum_j \psi^j(5\beta-\tau) \psi^j(\tau) \textrm{ for } 3\beta/2 \le \tau < 2\beta \textrm{ and } 3\beta \le \tau < 7\beta/2 \\
    2 \mu \sgn(9\beta-2\tau) \sum_j \psi^j(9\beta-\tau) \psi^j(\tau) \textrm{ for } 7\beta/2 \le \tau < 4\beta \textrm{ and } 5\beta \le \tau < 11\beta/2
    \end{cases}
    \label{eq:Renyi3SQ}
\end{split}
\end{equation}
Where $f_\tau = \Theta(\beta-\tau) + \Theta(\tau-2\beta) \Theta(3\beta-\tau) + \Theta(\tau-4\beta) \Theta(5\beta-\tau)$ and $\tilde f_\tau = \Theta(2\beta-\tau)\Theta(\tau-\beta)$ enforces the proper imaginary time evolution.

\begin{figure}
    \centering
    \includegraphics[width=0.75\linewidth]{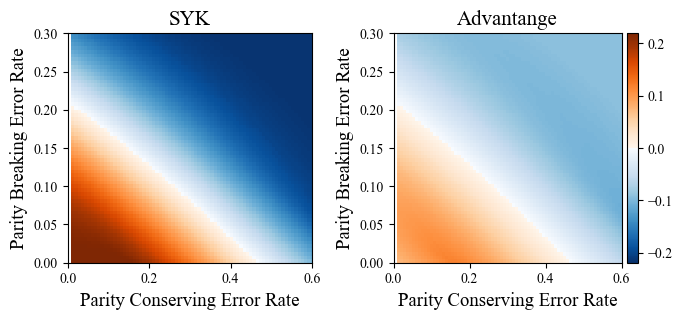}
    \includegraphics[width=0.75\linewidth]{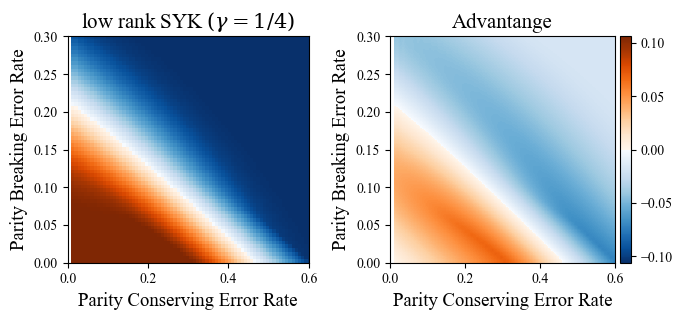}
    \caption{Renyi-3 Coherent information and relative advantage against no-encoding under both parity breaking and parity conserving decoherence for \textit{(Top)} the SYK model and \textit{(Bottom)} the low-rank SYK model at $\gam = 1/4$.
    }
    \label{fig:CohInfo_Renyi3both}
\end{figure}

The path integrals of Eq.\eqref{eq:Renyi3SQR} and Eq.\eqref{eq:Renyi3SQ} can be numerically found by solving for their large $N$ saddle-point.
Taking their difference, we find the Renyi-3 coherent information under parity breaking decoherence.
The results are given in Fig.\ref{fig:CohInfo_Renyi3both}c).
Just as with Renyi-2 coherent information, the Renyi-3 coherent information decreases continuously as a function of the parity breaking error rate $p$, and hence we do not have a error threshold against parity breaking errors.

\begin{figure}
    \centering
    \includegraphics[width=0.7\linewidth]{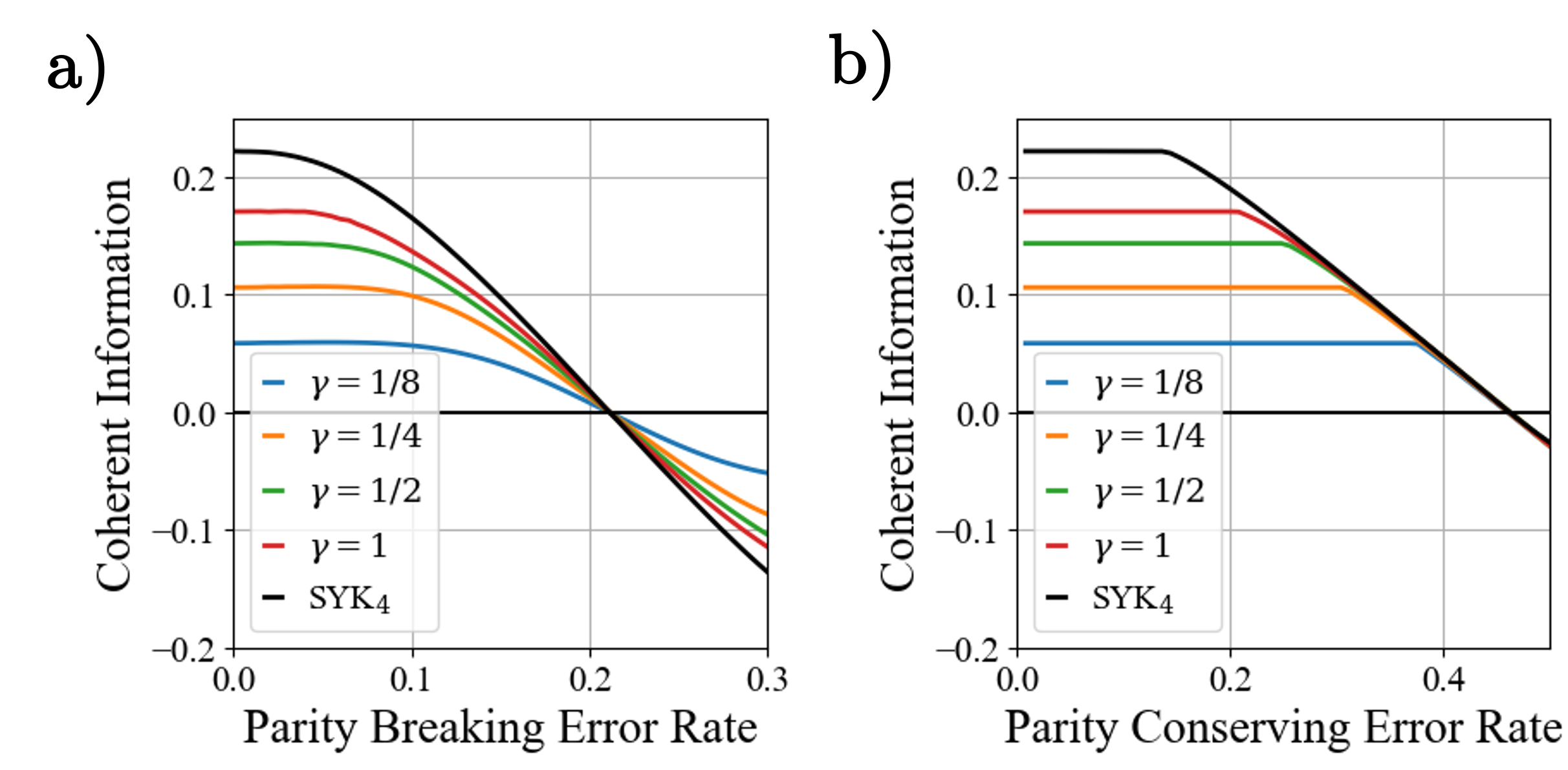}
    \caption{Renyi-3 coherent information against a) parity breaking decoherence and b) parity conserving decoherence. Just as with Renyi-2 coherent information, we see a threshold behavior of the coherent information under parity conserving decoherence.}
    \label{fig:CohInfo_Renyi3}
\end{figure}

In turn, we can utilize the Renyi-3 entropies against parity breaking decoherence to determine the Renyi-3 coherent information against parity conserving decoherence.
They are given as,
\begin{equation}
    \tilde{S}_{Q'R,Q'}^{(3)}(\cN^\inte_{\interp}) = {\min}_{\phi} \left\{\frac{3N}{2\interp} \phi^2 + \tilde S_{Q'R,Q'}^{(3)} \Big(\cN^\hop_{p(\phi)} \Big) \right\} \,.
\end{equation}
Taking their difference, we find the coherent information:
the results are given in Fig.\ref{fig:CohInfo_Renyi3}d).
As before, we find that there is a robust protection of the coherent information against parity conserving decoherence.

Last, we can consider the case of having both parity breaking and parity conserving noise at the same time.
The entanglement entropies of $\tilde S_{QR,Q}^{(3)}(\cN^\both)$ are given as,
\begin{equation}
    \tilde{S}_{Q'R,Q'}^{(3)}(\cN^\both_{p,\interp}) = {\min}_{\phi} \left\{\frac{3N}{2\interp} \phi^2 + \tilde S_{Q'R,Q'}^{(3)} \Big(\cN^\hop_{p(\phi+\phi_0)} \Big) \right\} \,, \textrm{ where } \phi_0 = \tanh^{-1} \frac{p_\hop}{1 - p_\hop} \,.
\end{equation}
The Renyi-3 coherent information against both types of noise are plotted in Fig.\ref{fig:CohInfo_Renyi3both} for the SYK model and for the low-rank SYK model at $\gamma = 1/4$.
We see qualitatively the same behaviors as the Renyi-2 coherent information.

\end{document}